\documentclass[12pt]{article}
\pdfoutput=1
\usepackage{subfigure}
\usepackage{amssymb,amsmath}
\usepackage{graphicx}
\usepackage{color}
\usepackage[colorlinks=true
,urlcolor=blue
,citecolor=blue
,linkcolor=blue
,pagecolor=blue
,linktocpage=true
,pdfproducer=medialab
]{hyperref}
\usepackage[a4paper,width=15.2cm]{geometry}
\makeatletter \renewcommand{\@dotsep}{10000} \makeatother

\begin{document}

\title{{ Neutralino-Sbottom Coannihilation in SU(5) }
}

\date{}
\maketitle
\vspace{-1.75cm}
\begin{center}
{\large
Ilia Gogoladze
\footnote{
Email: ilia@bartol.udel.edu. On leave of absence from:
Andronikashvili Institute of Physics, GAS, Tbilisi, Georgia.},
Shabbar Raza
\footnote{
Email: shabbar@udel.edu. On study leave from:
Department of Physics, FUUAST, Islamabad, Pakistan.}
and
Qaisar Shafi
\footnote{
Email: shafi@bartol.udel.edu. }
}

\vspace{0.75cm}

{\it
Bartol Research Institute, Department of Physics and Astronomy, \\
University of Delaware, Newark, DE 19716, USA 
}

\section*{Abstract}
\end{center}

We identify within the $SU(5)$ framework the minimum number of soft supersymmetry breaking parameters which can yield a bottom squrak (sbottom)
as the next to lightest supersymmetric particle. We focus in particular on the neutralino-sbottom coannihilation scenario
which gives rise to the desired neutralino dark matter relic density. We find solutions in which the sbottom mass is greater than or
of order $210\,{\rm GeV}$, while the gluino and the first two family squarks are heavier than $1\,{\rm TeV}$. Some benchmark points which can be tested
at the LHC are presented.

\thispagestyle{empty}
\setcounter{page}{0}
\newpage
\hrule
\bigskip
\tableofcontents
\bigskip\bigskip
\hrule


\section{Introduction}

In a recent paper \cite{Gogoladze:2011be}, the authors have shown that $b$-$\tau$ Yukawa coupling unification, realized in some well motivated $SU(5)$
and $SO(10)$ models, is compatible with the constrained minimal supersymmetric model (CMSSM) and with neutralino dark matter abundance
only if there exists neutralino-stop coannihilation. In order for this coannihilation scenario to be effective, the lighter stop must
be the next to lightest supersymmetric particle (NLSP), and it is quasi-degenerate in mass with the dark matter lightest supersymmetric particle (LSP) neutralino (to within 10-20$\%$ or so.)
A very recent analysis \cite{Adeel} of this neutralino-stop cooannihilation scenario shows that the ATLAS search for supersymmetry \cite{Aad:2011ib}, corresponding
to an integrated luminosity of 1 fb$^{-1}$, essentially rules out an NLSP stop mass below $140\,{\rm GeV}$. Future LHC searches
will no doubt provide far more stringent constraints on this scenario, or perhaps discover neutralino-stop coannihilation.
Another colored particle which can have mass of order $100\, {\rm GeV}$ and can still survive the recent ATLAS bounds \cite{arXiv:1103.4344} is the sbottom quark when
it is quasi-degenerate in mass with neutralino.

Motivated by these considerations we propose to investigate the neutralino-sbottom coannihilation scenario in this paper. The analysis in \cite{Gogoladze:2011be} shows
that this scenario requires a framework larger than the CMSSM with additional soft supersymmetry breaking (SSB) parameters. We have selected to
investigate neutralino-sbottom coannihilation in $SU(5)$, which naturally allows for the presence of additional parameters in accord with
minimal supergravity \cite{Chamseddine:1982jx}. We should note here that neutralino-sbottom coannihilation in $SU(5)$ has previously been explored in the references
listed in \cite{Profumo}. Our analysis, we believe, provides for the first time a comprehensive study of this scenario in $SU(5)$.

The outline for the rest of the paper is as follows. In section \ref{model} we briefly describe the model, list the $SU(5)$ inspired SSB parameters,
and the range of values employed in our scan. Section \ref{constraintsSection} describes the scanning procedure and the relevant experimental constraints that
we have employed. The results pertaining to neutralino-sbottom coannihilation are discussed in section \ref{sbottomCoan}, and our conclusions are summarized in section \ref{conclusions}.


\section{Soft Supersymmetry Breaking (SSB) Parameters in $SU(5)$ \label{model}}

We will search for neutralino-sbottom coannihilation in the $SU(5)$ framework and,
for simplicity, we assume that the SSB mass terms for  sfermions are family independent.
In $SU(5)$ the standard model (SM) fermions per family are allocated to the following representations: $\overline 5 \supset  (d^c, L)$ and $10\supset (Q, u^c, e^c)$, where in the brackets, we have employed standard notation for  the SM fermions.
It seems natural to consider two independent SSB scalar mass terms, at $M_{\rm G}$, $m_{\overline{5}}$ and $m_{10}$, for the matter multiplets.
The MSSM Higgs doublets belong to $5 (H_u)$ and $\overline 5 (H_{d})$ representations of $SU(5)$, to which we assign two independent SSB mass terms, $m_{H_u}$ and $m_{H_d}$.

The minimal $SU(5)$ model predicts $b$-$\tau$ Yukawa coupling unification at $M_{\rm G}$ from the interaction ${\overline{5}}.10.{\overline{5}_{H_d}}$, and its
 LHC implications have been discussed in \cite{Gogoladze:2011be}. It was shown in \cite{Gogoladze:2008dk} that $b$-$\tau$ Yukawa coupling unification can be
 relaxed by including either non-renormalizable interactions in the theory, or by employing a more complicated Higgs sector, beyond the minimal one. To implement
neutralino-sbottom coannihilation, we find it helpful not to require $b$-$\tau$ Yukawa unification. We also invoke non-universal soft  trilinear terms.
In particular,  we assume that at $M_{\rm G}$, there are two independent parameters $A_b(=A_{\tau})$ and $A_t$, where the equality $A_b=A_{\tau}$
keeps the number of SSB parameters to a minimum. To summarize, we consider the following SSB terms for this study:
\begin{align}
M_{1/2}, m_{10},  m_{\overline{5}}, m_{H_u},m_{H_d}, \tan\beta, A_{t}, A_{b}(=A_{\tau}) \,\, {\rm{and}} \,\,   sign(\mu).
\label{eq1}
\end{align}
We shall see that this is the minimal set of independent MSSM terms which allows one to implement neutralino-sbottom coannihilation in $SU(5)$. We will
set $\mu>0$ in this paper.

\section{Phenomenological constraints and scanning procedure\label{constraintsSection}}

We employ the ISAJET~7.80 package~\cite{ISAJET}  to perform random
scans over the parameters listed in Eq.(\ref{eq1}). In this
package, the weak scale values of gauge and third generation Yukawa
couplings are evolved to $M_{\rm G}$ via the MSSM renormalization
group equations (RGEs) in the $\overline{DR}$ regularization scheme.
We do not strictly enforce the unification condition $g_3=g_1=g_2$ at $M_{\rm
G}$, since a few percent deviation from unification can be
assigned to unknown GUT-scale threshold
corrections~\cite{Hisano:1992jj}.
The difference between $g_1(=g_2)$ and $g_3$ at $M_{G}$ is no
worse than $4\%$.

The various boundary conditions are imposed at
$M_{\rm G}$ and all the SSB
parameters, along with the gauge and Yukawa couplings, are evolved
back to the weak scale $M_{\rm Z}$.
In the evaluation of Yukawa couplings the SUSY threshold
corrections~\cite{Pierce:1996zz} are taken into account at the
common scale $M_{\rm SUSY}= \sqrt{m_{{\tilde t}_L}m_{{\tilde t}_R}}$, where
$m_{{\tilde t}_L}$ and $m_{{\tilde t}_R}$ are the third generation left and right
handed stop quarks. The entire
parameter set is iteratively run between $M_{\rm Z}$ and $M_{\rm
G}$ using the full 2-loop RGEs until a stable solution is
obtained. To better account for leading-log corrections, one-loop
step-beta functions are adopted for gauge and Yukawa couplings, and
the SSB parameters $m_i$ are extracted from RGEs at multiple scales
$m_i=m_i(m_i)$. The RGE-improved 1-loop effective potential is
minimized at an optimized scale $M_{\rm SUSY}$, which effectively
accounts for the leading 2-loop corrections. Full 1-loop radiative
corrections are incorporated for all sparticle masses.

The requirement of  radiative electroweak symmetry breaking  puts an important theoretical
constraint on the parameter space. Another important constraint
comes from limits on the cosmological abundance of stable charged
particles~\cite{Nakamura:2010zzi}. This excludes regions in the parameter space
where charged SUSY particles, such as ${\tilde \tau}_1$ or ${\tilde t}_1$, become
the LSP. We keep only those
solutions where the lightest neutralino is the LSP which, in most cases, saturates
the WMAP (Wilkinson Microwave Anisotropy Probe) dark matter relic abundance bound.
Neutralino-sbottom coannihilation plays an important role in realizing the desired
LSP relic abundance.

We have performed random scans for the following parameter range:
\begin{align}
0\leq   m_{5}, m_{H_d} \leq 5\, \rm{TeV} \nonumber \\
0\leq  m_{10}, m_{H_u},  \leq 10\, \rm{TeV} \nonumber \\
0 \leq M_{1/2}  \leq 2\,\rm {TeV} \nonumber \\
-15\rm{TeV} \leq A_t  \leq 15\, \rm{TeV} \nonumber \\
-15\rm{TeV} \leq A_b=A_\tau  \leq 30\, \rm{TeV} \nonumber \\
1.1\leq \tan\beta \leq 60 \nonumber \\
\mu>0
 \label{parameterRange}
\end{align}
where $m_t = 173.3\pm1.1\, {\rm GeV}$ \cite{:2009ec}  is the top quark pole mass.
We use $m_b(m_Z)=2.83$ GeV which is hard-coded into ISAJET. The various boundary conditions in Eq.(\ref{parameterRange}) are implemented through
the scanning procedure. We found, for instance, that $m_{\overline{5}}$ has to be smaller than $m_{10}$ to realize the neutralino-sbottom coannihilation solutions.

In scanning the parameter space, we employ the Metropolis-Hastings
algorithm as described in \cite{Belanger:2009ti}. All of the
collected data points satisfy
the requirement of radiative electroweak symmetry breaking (REWSB),
with the lightest neutralino in each case being the LSP.
After collecting the data, we impose
the mass bounds on all the particles~\cite{Nakamura:2010zzi} and use the
IsaTools package~\cite{Baer:2002fv}
to implement the following phenomenological constraints on points that
have sbottom coannihilation solution:
\begin{table}[h]\centering
\begin{tabular}{rlc}
$m_h~{\rm (lightest~Higgs~mass)} $&$ \geq\, 114.4~{\rm GeV}$                    &  \cite{Schael:2006cr} \\
$BR(B_s \rightarrow \mu^+ \mu^-) $&$ <\, 1.1 \times 10^{-8}$                     &   \cite{CMS_plus_LHCb}      \\
$2.85 \times 10^{-4} \leq BR(b \rightarrow s \gamma) $&$ \leq\, 4.24 \times 10^{-4} \; (2\sigma)$ &   \cite{Barberio:2007cr}  \\
$0.15 \leq \frac{BR(B_u\rightarrow \tau \nu_{\tau})_{\rm MSSM}}{BR(B_u\rightarrow \tau \nu_{\tau})_{\rm SM}}$&$ \leq\, 2.41 \; (3\sigma)$ &   \cite{Barberio:2008fa}  \\
$\Omega_{\rm CDM}h^2 $&$ =\, 0.111^{+0.028}_{-0.037} \;(5\sigma)$               &  \cite{Komatsu:2008hk}
\end{tabular}
\end{table}

{\noindent}As far as  the muon anomalous  magnetic moment   is concerned, we only require that the model does no worse than the  SM.

\section{Sbottom Coannihilation \label{sbottomCoan}}

\begin{figure}
\centering
\subfiguretopcaptrue
\subfigure{
\includegraphics[width=7.cm]{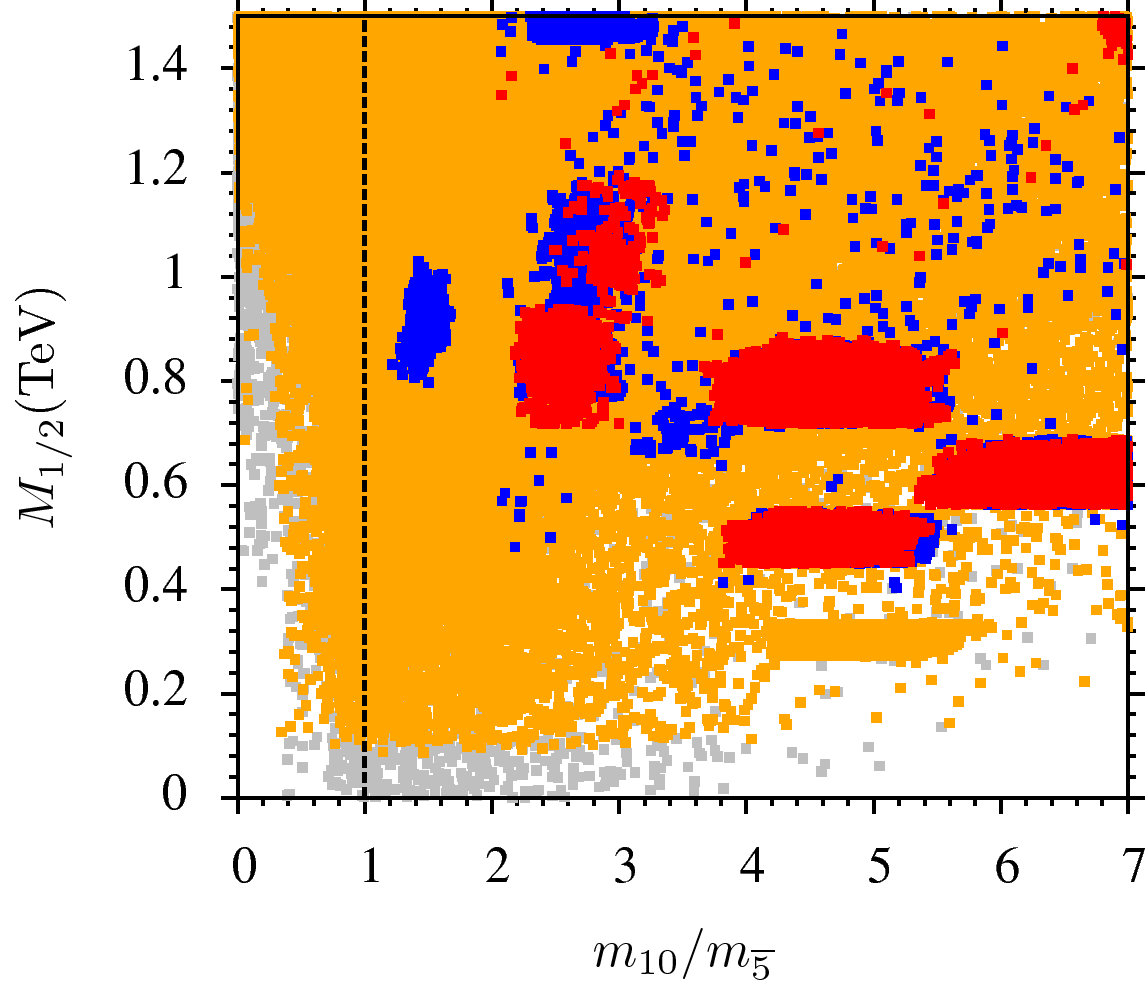}
}
\subfigure{
\includegraphics[width=7.cm]{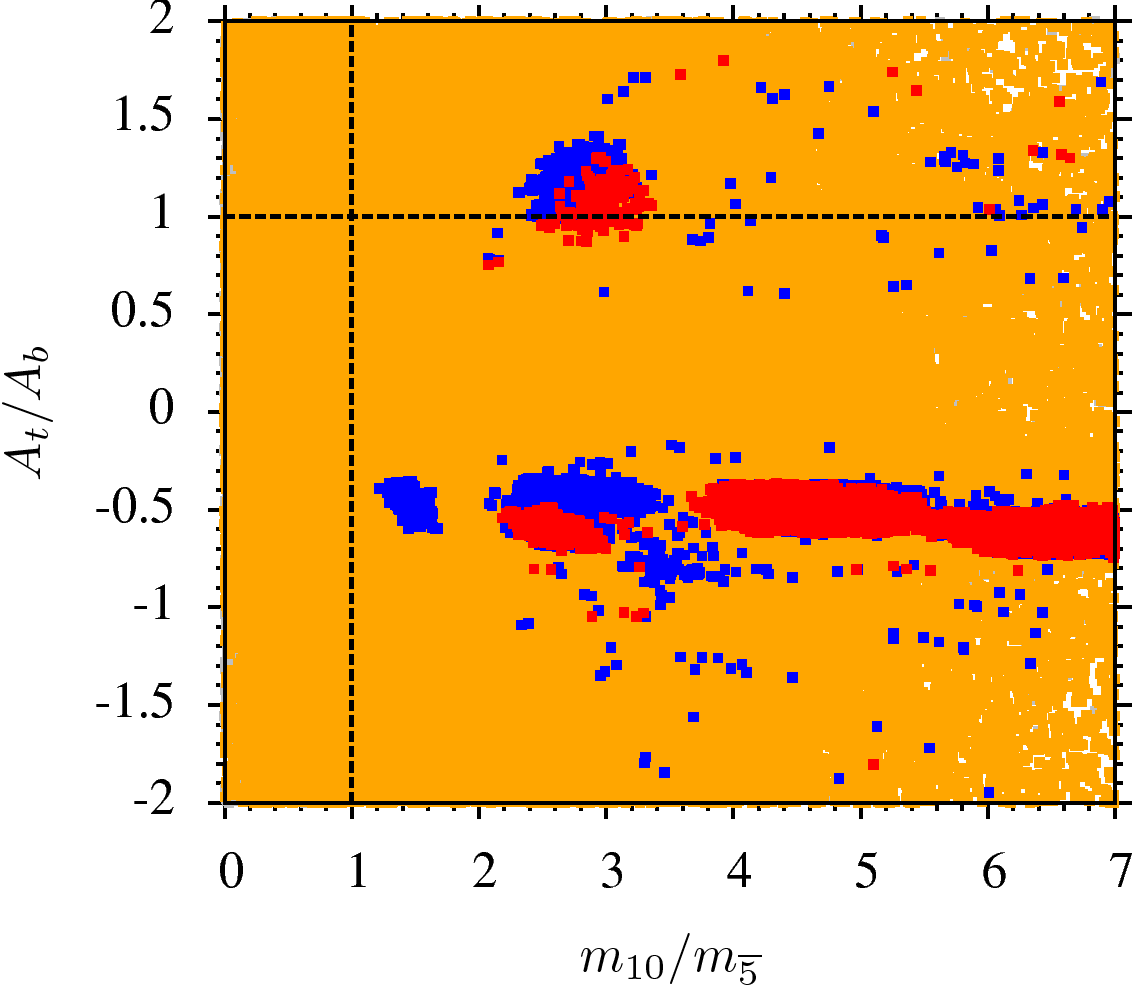}
}
\subfigure{
\includegraphics[width=7.cm]{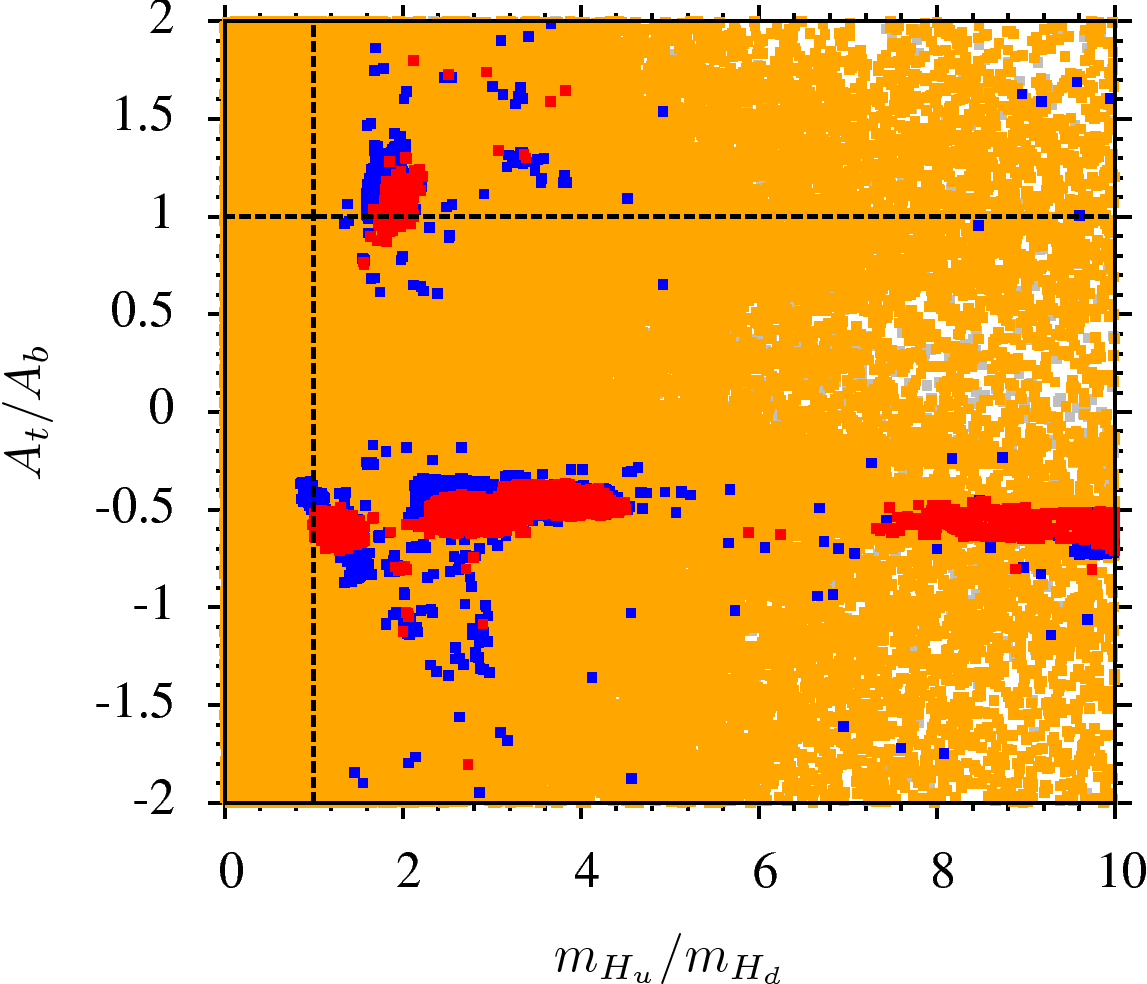}
}
\subfigure{
\includegraphics[,width=7.cm]{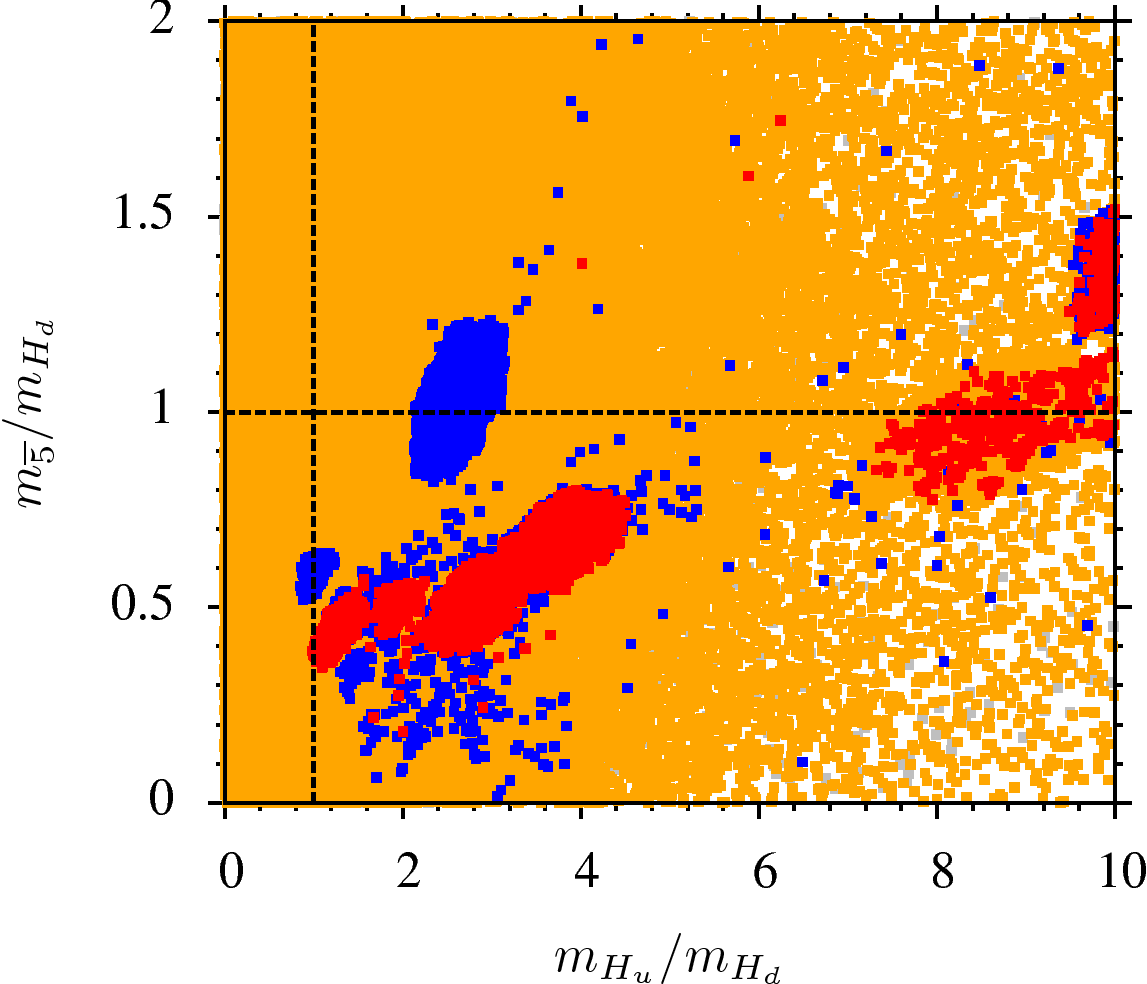}
}
\caption{Plots in the $M_{1/2}$ - $m_{10}/m_{\overline {5}}$,
 $A_t/A_b$ - $m_{10}/m_{\overline {5}}$,  $A_t/A_b$  - $m_{H_u}/m_{H_d}$,  $m_{\overline 5}/m_{H_d}$ - $m_{H_u}/m_{H_d}$
planes. Gray points are consistent with REWSB, neutralino  LSP.
The orange points satisfy, in addition, the particle mass
bounds, constraints from $BR(B_s\rightarrow \mu^+ \mu^-)$,  $BR(B_u\rightarrow \tau \nu_{\tau})$  and $BR(b\rightarrow s \gamma)$.
The blue points form a subset of orange points that shows sbottom quark as an NLSP. The red points represent a subset of blue points that corresponds to
the neutralino-sbottom  coannihilation scenario. The vertical and horizontal dashed  lines show equality of parameters along x and y-axes respectively.}
\label{Fig1}
\end{figure}

To realize the desired LSP dark matter relic abundance through  neutralino-sbottom coannihilation, the
mass difference between these two should be less than or of order 20$\%$ \cite{Profumo}.
In Figure \ref{Fig1} we present our results in the $M_{1/2}$ - $m_{10}/m_{\overline {5}}$,
 $A_t/A_b$ - $m_{10}/m_{\overline {5}}$,  $A_t/A_b$  - $m_{H_u}/m_{H_d}$,  $m_{\overline 5}/m_{H_d}$ - $m_{H_u}/m_{H_d}$
planes, with the parameter values all defined at $M_{\rm G}$. Gray points are consistent with REWSB and LSP neutralino.
The orange  points satisfy, in addition,  the particle mass
bounds, constraints from $BR(B_s\rightarrow \mu^+ \mu^-)$,  $BR(B_u\rightarrow \tau \nu_{\tau})$  and $BR(b\rightarrow s \gamma)$.
The blue points form a subset of orange points and correspond to NLSP sbottom, but which is not closely degenerate in mass with the neutralino.
The red points form a subset of blue points and represent neutralino-sbottom coannihilation, the scenario we are really after $!$
The vertical and the horizontal dashed lines show equality of parameters along the x and y-axes respectively.

In order to explain our findings we consider the one loop renormalization group equations for the third generation squarks and sleptons \cite{Martin:1997ns}
\begin{eqnarray}
16\pi^2 \frac{d}{dt} m_{\tilde Q_3}^2 &=&
X_t +X_b-\frac{32}{3} g_3^2 |M_3|^2 -6 g_2^2 |M_2|^2 -\frac{2}{5} g_1^2 |M_1|^2
+ \frac{1}{5} g^{2}_1 S,
\label{mQ3}
\\
16\pi^2 \frac{d}{dt} m_{\tilde u_3^c}^2 &=&
2 X_t -\frac{32}{3} g_3^2 |M_3|^2 - \frac{32}{15} g_1^2 |M_1|^2
- \frac{4}{5} g^{2}_1 S,
\label{mU3}
\\
16\pi^2 \frac{d}{dt} m_{\tilde d_3^c}^2  &=&
2 X_b - \frac{32}{3} g_3^2 |M_3|^2 - \frac{8}{15} g_1^2|M_1|^2
+ \frac{2}{5} g^{2}_1 S ,
\label{mD3}
\\
16\pi^2 \frac{d}{dt} m_{\tilde L_3}^2  &=&
X_\tau  - 6 g_2^2 |M_2|^2 - \frac{6}{5} g_1^2 |M_1|^2 - \frac{3}{5} g^{2}_1 S,
\label{ml3}
\\
16\pi^2 \frac{d}{dt} m_{\tilde e_3^c}^2 &=&
2 X_\tau - \frac{24}{5} g_1^2 |M_1|^2  + \frac{6}{5} g^{2}_1 S.
\label{me3}
\end{eqnarray}
Here
\begin{eqnarray}
X_t  &=  & 2 |y_t|^2 (m_{H_u}^2 + m_{\tilde Q_3}^2 + m_{\tilde u_3^c}^2) +2 |A_t|^2,
\label{xt}
\\
X_b  &= & 2 |y_b|^2 (m_{H_d}^2 + m_{\tilde Q_3}^2 + m_{\tilde d_3^c}^2) +2 |A_b|^2,
\label{xb}
\\
X_\tau  &= &  2 |y_\tau|^2 (m_{H_d}^2 + m_{\tilde L_3}^2 + m_{\tilde e_3^c}^2)
+ 2 |A_\tau|^2,
\label{xtau}
\\
S &\equiv & {\rm Tr}[Y_j m^2_{\phi_j}] =
m_{H_u}^2 - m_{H_d}^2 + {\rm Tr}[
{ m^2_{\tilde Q}} - { m^2_{\tilde L}} - 2 { m^2_{{\tilde u}^c}}
+ { m^2_{{\tilde d}^c}} + { m^2_{{\tilde e}^c}}] ,
\label{eq:defS}
\end{eqnarray}
and ${\tilde Q_3}$, ${\tilde u_3^c}$, $\tilde d_3^c$, ${\tilde L_3}$, ${\tilde e_3^c}$ denote the third generation squarks and sleptons.
Also, $g_i$ and $M_i$  ($i=1,2,3$) denote the gauge couplings and gaugino masses
for $U(1)_Y$, $SU(2)_L$ and $SU(3)_C$, and $y_j$, $A_j$ ($j=t, b, \tau$) are the third family Yukawa couplings and trilinear scalar SSB couplings, respectively.

In the $M_{1/2}$ - $m_{10/}m_{\overline {5}}$ plane of Figure \ref{Fig1}, we show that in order to have sbottom NLSP, the ratio
$m_{10}/m_{\overline 5}$  should be greater than 1.4 (blue points). If we require successful neutralino-sbottom coannihilation,
the ratio $m_{10}/m_{\overline 5}>2$ (red points).
In general, there is a two step process for realizing NLSP sbottom. First we consider the RGE effects, and in the second step we search
for cancellation in the $2\times 2 $ sbottom mass matrix between the diagonal and off diagonal ($-m_b(A_b+\mu\tan\beta)$) entries.

Let us see why it is not possible to have sbottom NLSP starting from universal
sfermion masses ($m_{\overline {5}}=m_{10}$).  First, we note, that it is difficult to make the sbottom lighter than the stop.
Eqs.(\ref{mQ3})-(\ref{eq:defS}) show that the gluino  loop  contribution raises the squark  masses (we ignore the hypercharge contribution
because it is sufficiently small), while the Yukawa and SSB trilinear couplings tend to lower them. The left handed squarks are heavier than the
right handed ones due to the $SU(2)_L$ contribution. From the  QCD point of
view the sbottom and stop masses$^2$ renormalize identically. To split them, the loop corrections involving Yukawa couplings must play a role.
Since the top Yukawa coupling, for most $\tan\beta$ values, is larger than the bottom Yukawa coupling, it is hard to make $m_{\tilde{b}^c}^2<m_{\tilde{t^c}}^2$ through RG running. If the soft trilinear couplings $A_t$ and $A_b $ are independent of each other, one could make one of the sbottom mass eigenvalues lighter than the stop. But in this case we need to make sure that the eigenvalues of the stau mass$^2$ matrix remain positive, after suitable
cancellation in the sbottom mass$^2$ matrix is accomplished.
 It turns out, as we show in  $A_t/A_b$ - $m_{10}/m_{\overline {5}}$ plane (Figure \ref{Fig1}), that $m_{10}=m_{\overline {5}}$ and $ A_t=A_b$ are not compatible with neutralino-sbottom coannihilation. It is possible to achieve neutralino-sbottom coannihilation with $A_t=A_b=A_{\tau}$,
provided $m_{10}/m_{\overline {5}}>2$. The ratio $m_{10}/m_{\overline {5}}$ takes its minimal value around $1.4$, which corresponds to $A_t/A_b< - 0.5$.
Note that  for $M_{1/2}>m_{0}$ where $m_0$ is the universal SSB mass term for sfermions at  $M_{\rm G}$, it is difficult to make the sbottom lighter than the right handed stau because of QCD corrections from the gluino loop. Combining these two observations, it is clear that we need
$m_{10}>m_{\overline 5}$ for any given value of $M_{1/2}$.

  In the $A_t/A_b$  - $m_{H_u}/m_{H_d}$ plane in Figure \ref{Fig1} we see two viable neutralino-sbottom coannihilation regions which have $A_t/A_b=1$ or $m_{H_u}/m_{H_d}=1$. However, we also see that it is not possible to have such solutions with both these ratios simultaneously equal to unity.

In minimal $SU(5)$ the MSSM  down type Higgs field resides in the $\overline 5$ representation, and so it is interesting to see whether or not
 neutralino-sbottom coannihilation or sbottom NLSP is possible by setting $m_{\overline 5}=m_{H_d}$. In the $m_{\overline 5}/m_{H_d}$ - $m_{H_u}/m_{H_d}$ plane
with $m_{\overline 5}=m_{H_d}$, neutralino-sbottom coannihilation requires $m_{H_u}/m_{H_d}\geq 8$ (red points), while sbottom NLSP only needs $m_{H_u}/m_{H_d}\geq 2$ (blue points).

\begin{figure}
\centering
\subfiguretopcaptrue
\subfigure{
\includegraphics[totalheight=5.5cm,width=7.cm]{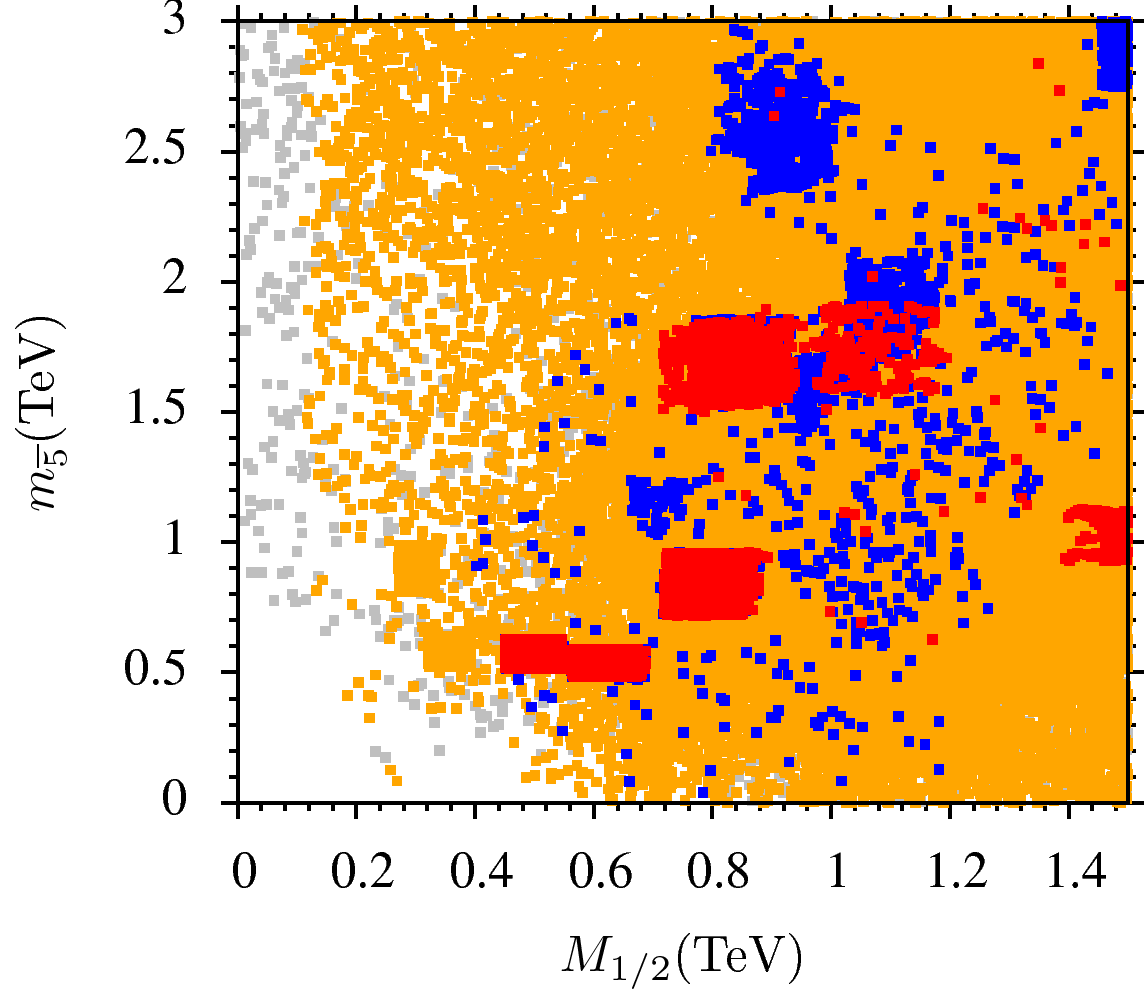}
}
\subfigure{
\includegraphics[totalheight=5.5cm,width=7.cm]{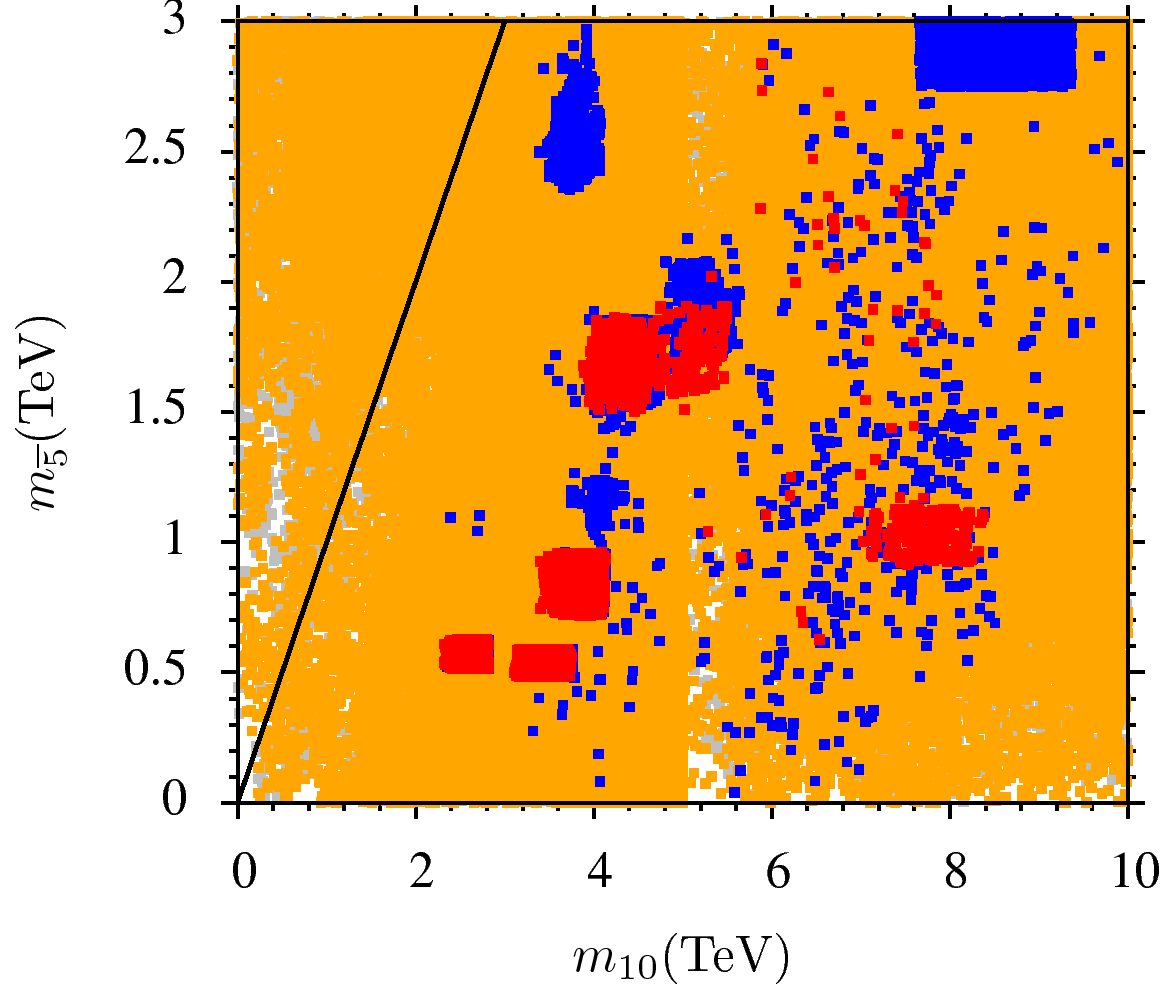}
}
\subfigure{
\includegraphics[totalheight=5.5cm,width=7.cm]{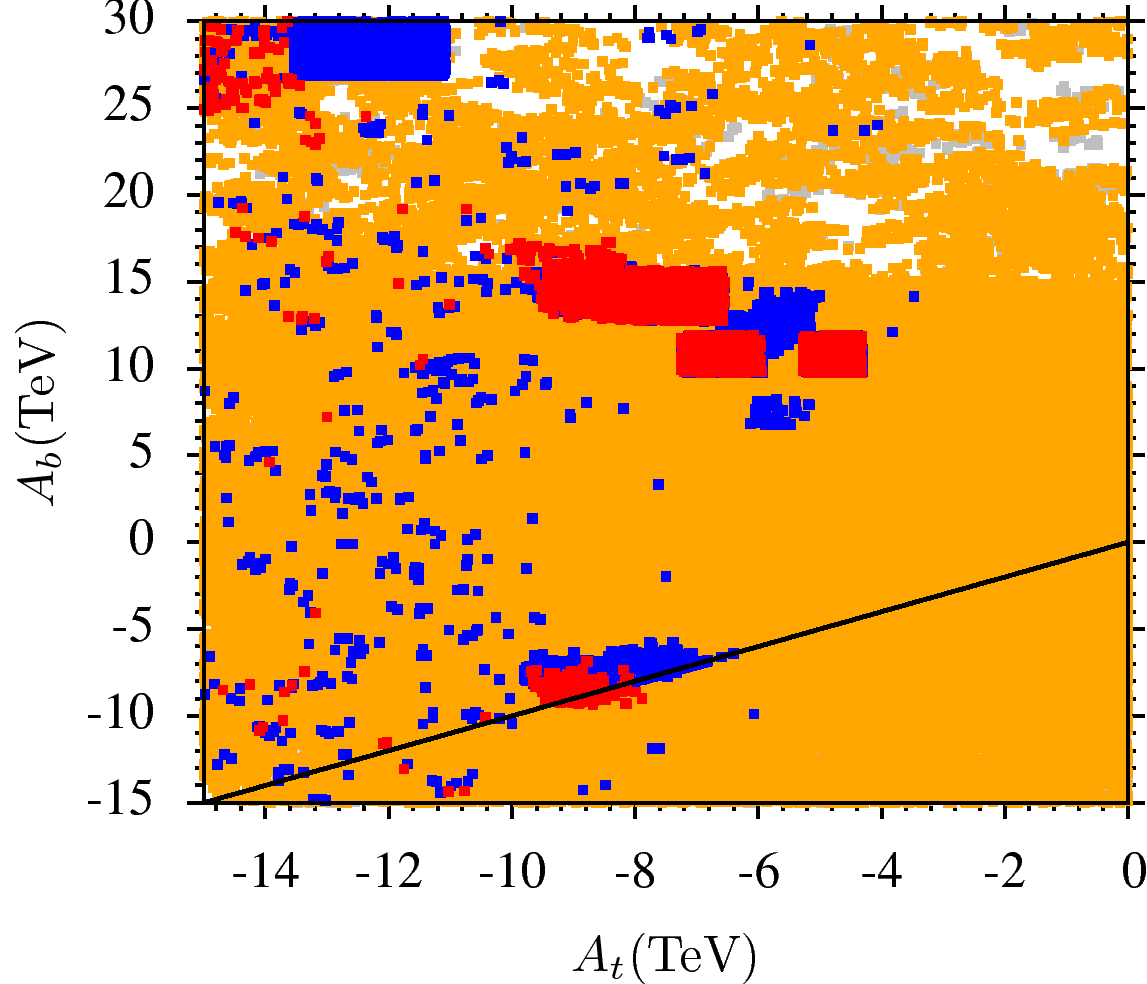}
}
\subfigure{
\includegraphics[totalheight=5.5cm,width=7.cm]{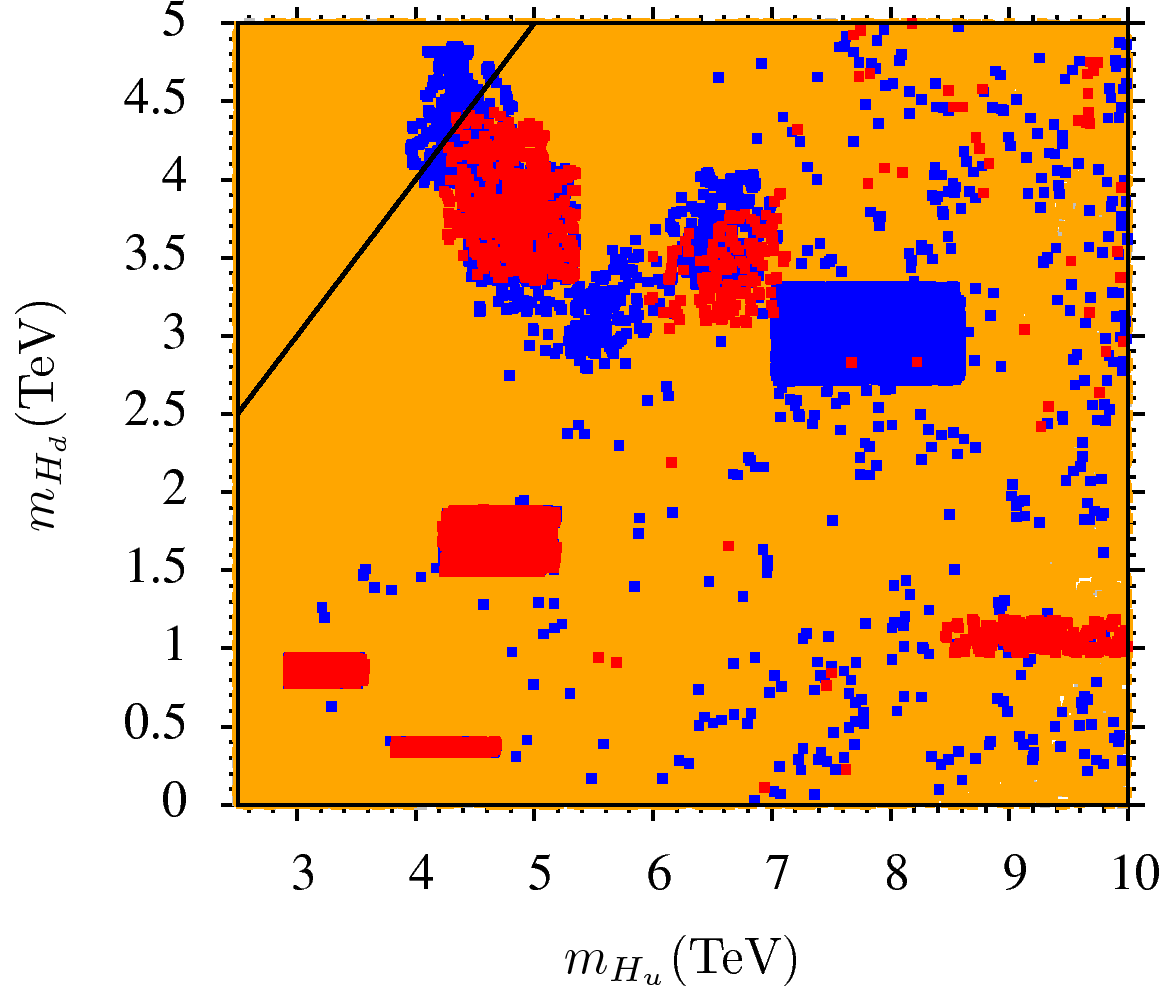}
}
\subfigure{
\includegraphics[totalheight=5.5cm,width=7.cm]{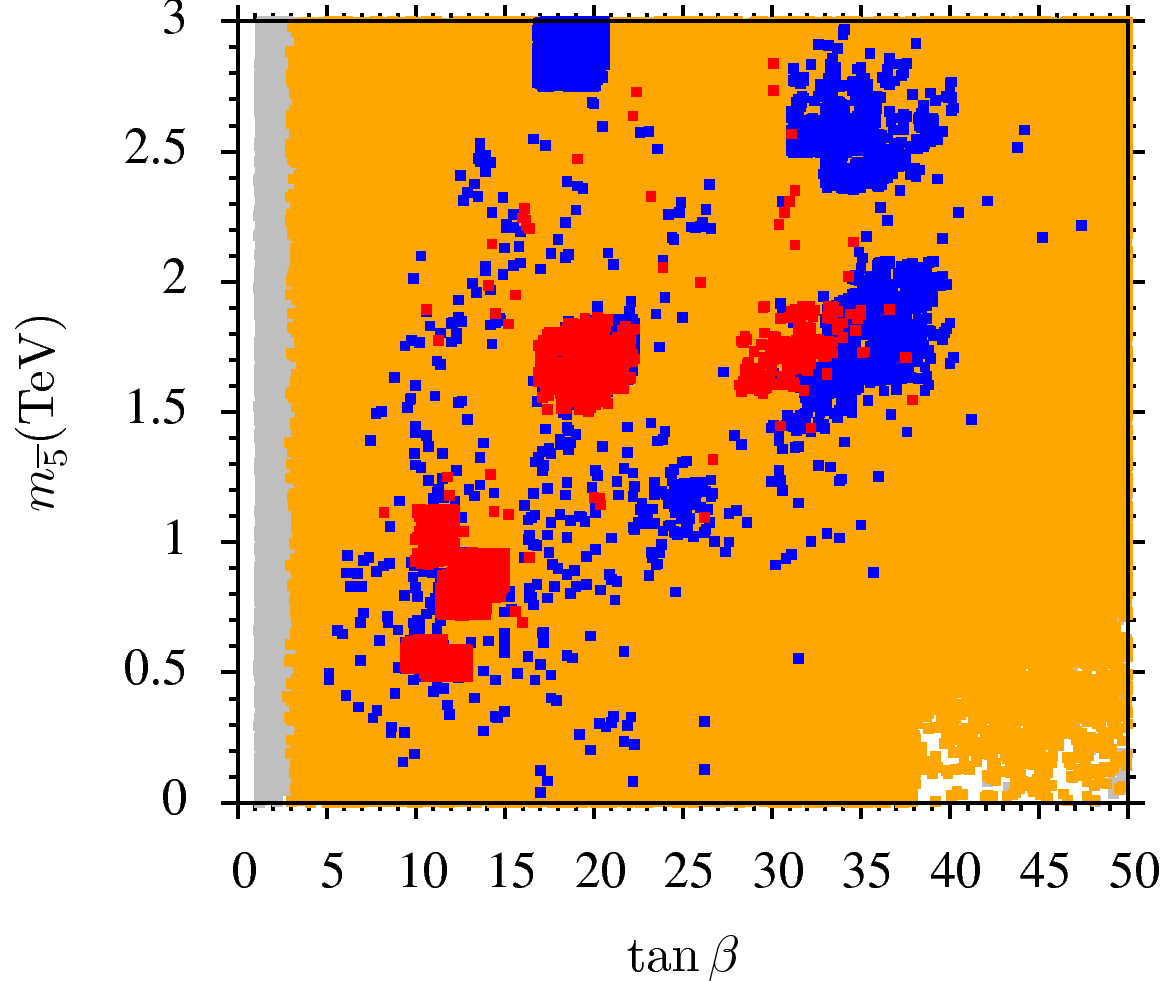}
}
\subfigure{
\includegraphics[totalheight=5.5cm,width=7.cm]{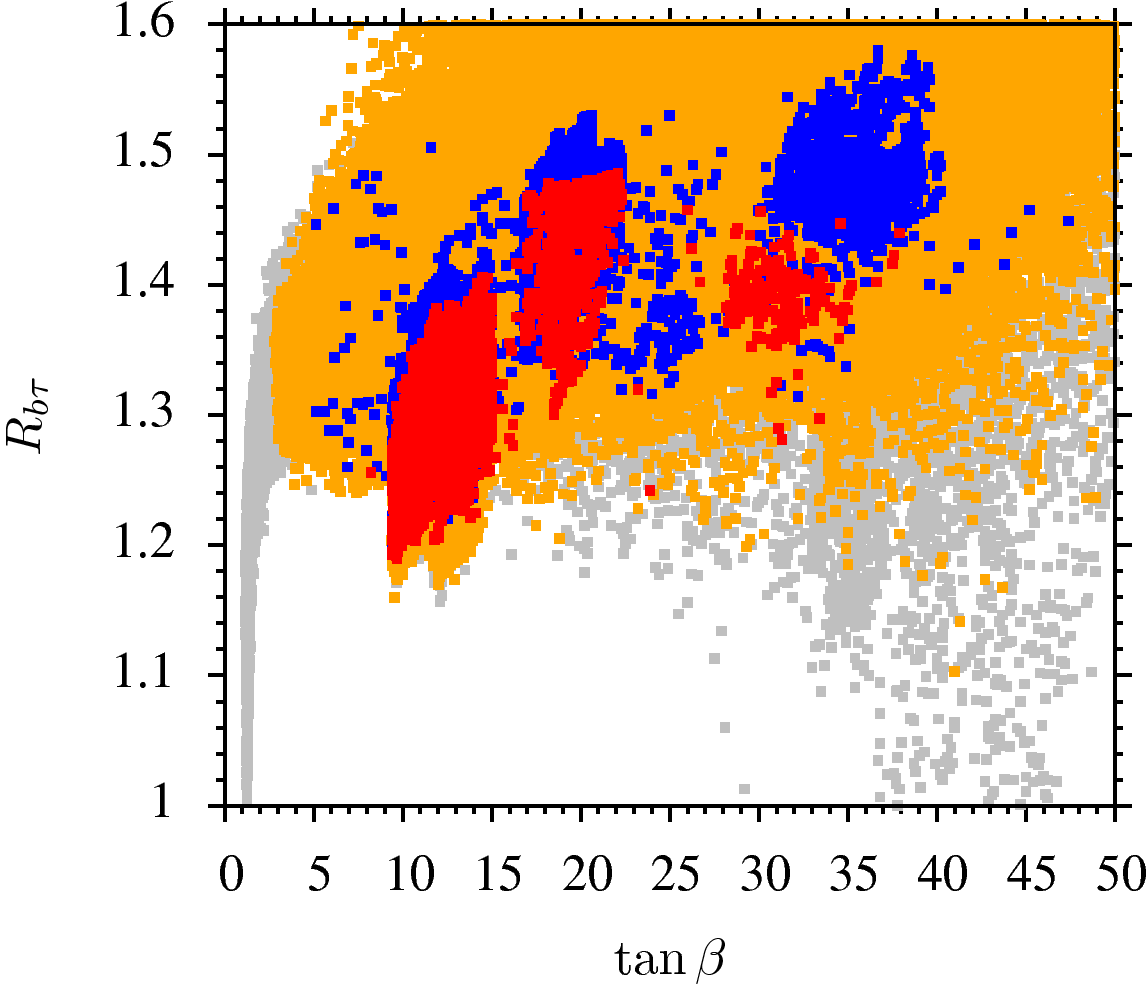}
}
\caption{Plots in the $m_{1/2}$ - $ m_{\overline {5}}$,  $m_{10}$ - $m_{\overline {5}}$,
 $A_t$ - $A_b$, $m_{H_d}$ - $m_{H_u}$,
 $\tan\beta$ - $ m_{\overline {5}}$ and $R_{b\tau}$ -  $\tan\beta$
planes.
Color coding
same as in Figure~\ref{Fig1}.
Lines in black have unit slopes with x and y coordinates equal}.
\label{Fig2}
\end{figure}

In Figure \ref{Fig2} we present results
 in the $m_{\overline 5}$ - $M_{1/2}$,  $m_{\overline 5}$ - $m_{10}$,
 $A_t$ -  $A_b$, $m_{H_d}$ - $m_{H_u}$, $\tan\beta$ - $ m_{\overline {5}}$ and $R_{b\tau}$ -  $\tan\beta$
planes which display the values of the SSB parameters required to yield sbottom NLSP/ coannihilation.
Gray points are consistent with REWSB and neutralino  LSP.
The orange points satisfy, in addition,  the particle mass
bounds, constraints from $BR(B_s\rightarrow \mu^+ \mu^-)$,  $BR(B_u\rightarrow \tau \nu_{\tau})$  and $BR(b\rightarrow s \gamma)$.
The blue points form a subset of orange points and correspond to sbottom NLSP. The red points represent a subset of the blue points
and represent the neutralino-sbottom  coannihilation scenario.

As we can see from the $m_{\overline 5}$ - $M_{1/2}$ panel, neutralino-sbottom coannihilation prefers relatively heavy gauginos
($M_{1/2} > 400$ GeV), while $m_{\overline {5}}$ can be as light as $500\, {\rm GeV}$. For  $m_{\overline {5}}$ lighter than 500 GeV, it is
difficult to separate the right handed sbottom mass from the left handed tau slepton.  In the $m_{\overline {5}}$ - $m_{10}$ plane, we see that
$m_{10}$ should be larger than 2 TeV in order to have neutralino-sbottom coannihilation.

As noted above, the  RG running is not sufficient to realize an NLSP sbottom,
 and some cancellation is required in addition. Consider the plot in $A_t$-$A_b$ plane at SUSY scale where we see that neutralino-sbottom coannihilation
solutions mostly require $|A_b|> |A_t|$, where $|A_b|$ can be $O(10-15){\rm TeV}$. The off-diagonal entries $m_b(A_b +\mu \tan\beta)$
for the sbottom quark mass matrix are of comparable
magnitude to the diagonal entries and, as a result, we can realize neutralino-sbottom coannihilation.

The plot in $m_{H_d}$-$m_{H_u}$ (at $M_{\rm G}$) plane shows
that neutralino-sbottom coannihilation requires $m_{H_u}>m_{H_d}$.
This can be understood by considering the one-loop RGEs for $m_{H_u}^2$
and $m_{H_d}^2$:
\begin{eqnarray}
16 \pi^2 \frac{d}{dt} m_{H_u}^2 &=&
3 X_t - 6 g_2^2 |M_2|^2 - \frac{6}{5} g_1^2 |M_1|^2 + \frac{3}{5} g^{2}_1 S,
\label{mhurge}
\\
16\pi^2 \frac{d}{dt} m_{H_d}^2 &=&
3 X_b + X_\tau - 6 g_2^2 |M_2|^2 - \frac{6}{5} g_1^2 |M_1|^2 - \frac{3}{5}
g^{2}_1 S.
\label{mhdrge}
\end{eqnarray}
Combining with Eqs. (\ref {xt})-(\ref{xtau}) we see that  the RG evolution of $m_{H_u}^2$ and  $m_{H_d}^2$ depend on $y_t$, $y_b$, $|A_b|$ and $|A_t|$.
Our conclusion  from Figure \ref{Fig1}  is  that neutralino-sbottom coannihilation  solutions require in most cases  $|A_t|<|A_b|$, which means
that $m_{H_d}^2$ renormalizes (decreases)  more than  $m_{H_u}^2$.
In order to have $-m^2_{H_d} > -m^2_{H_u}$ at low scale, we need to start in most cases with $m^2_{H_d} < m^2_{H_u}$ at $M_{\rm G}$.

We note from $m_{\overline 5}$ -  $\tan\beta$  plane  that sbottom coannihilation prefers small to moderately large values for  $\tan\beta$ ($7<tan\beta<40$). In this region the
top Yukawa coupling reaches its minimum value at $M_{\rm G}$.

In order to quantify $b$-$\tau$ Yukawa unification, we define the quantity $R_{b\tau}$ as
\begin{align}
R_{b\tau}=\frac{ {\rm max}(y_b,y_{\tau})} { {\rm min} (y_b,y_{\tau})}.
\end{align}
The $R_{b\tau}$-$\tan\beta$ plot in Figure \ref{Fig2} shows that it is  difficult to reconcile $b$-$\tau$ Yukawa unification
with the neutralino-sbottom coannihilation scenario. For the parameter range given in Eq.(\ref{parameterRange}), $b$-$\tau$ Yukawa unification is
at the level of 20$\%$ or so, which agrees with the results in ref. \cite{Gogoladze:2008dk}.

\begin{figure}
\centering
\subfiguretopcaptrue
\subfigure{
\includegraphics[totalheight=5.5cm,width=7.cm]{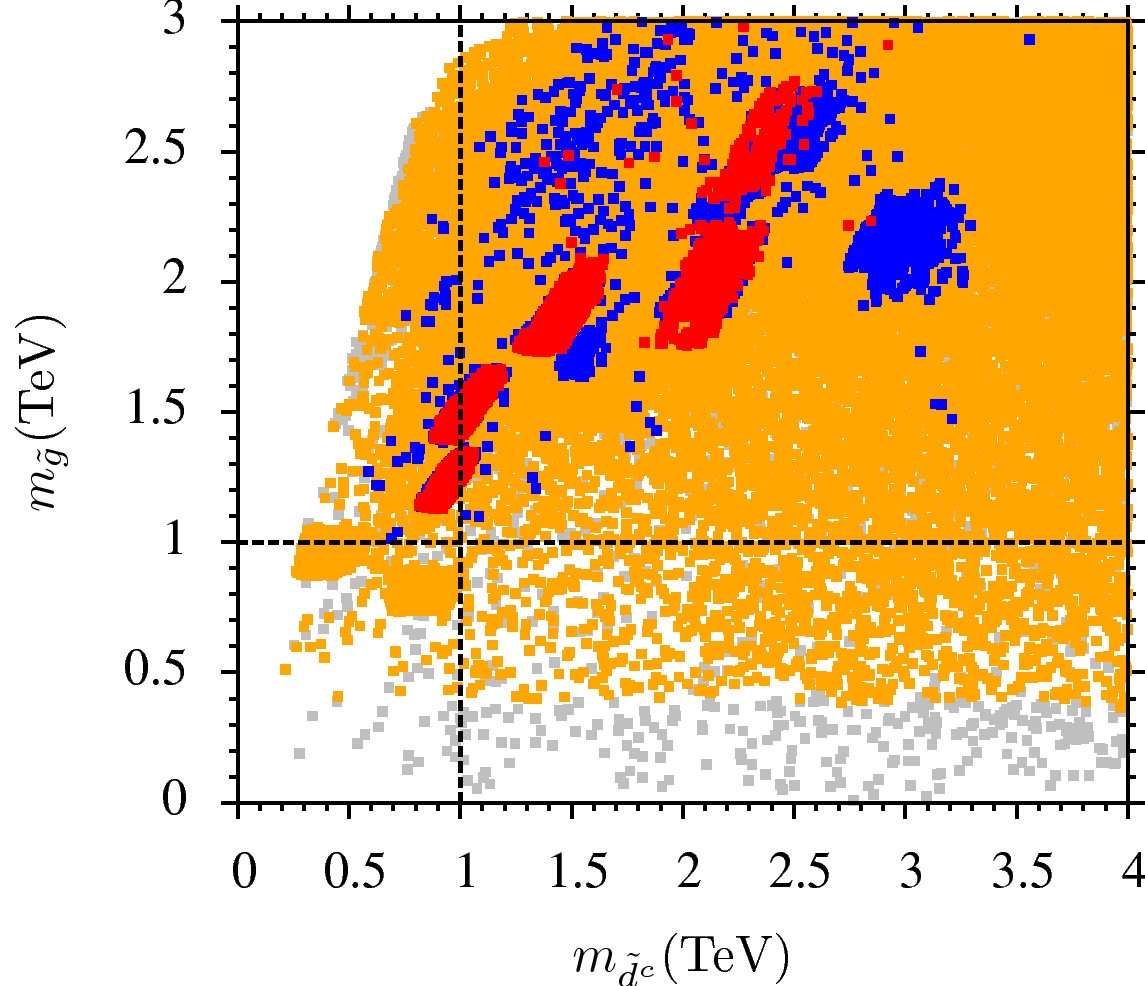}
}
\subfigure{
\includegraphics[totalheight=5.5cm,width=7.cm]{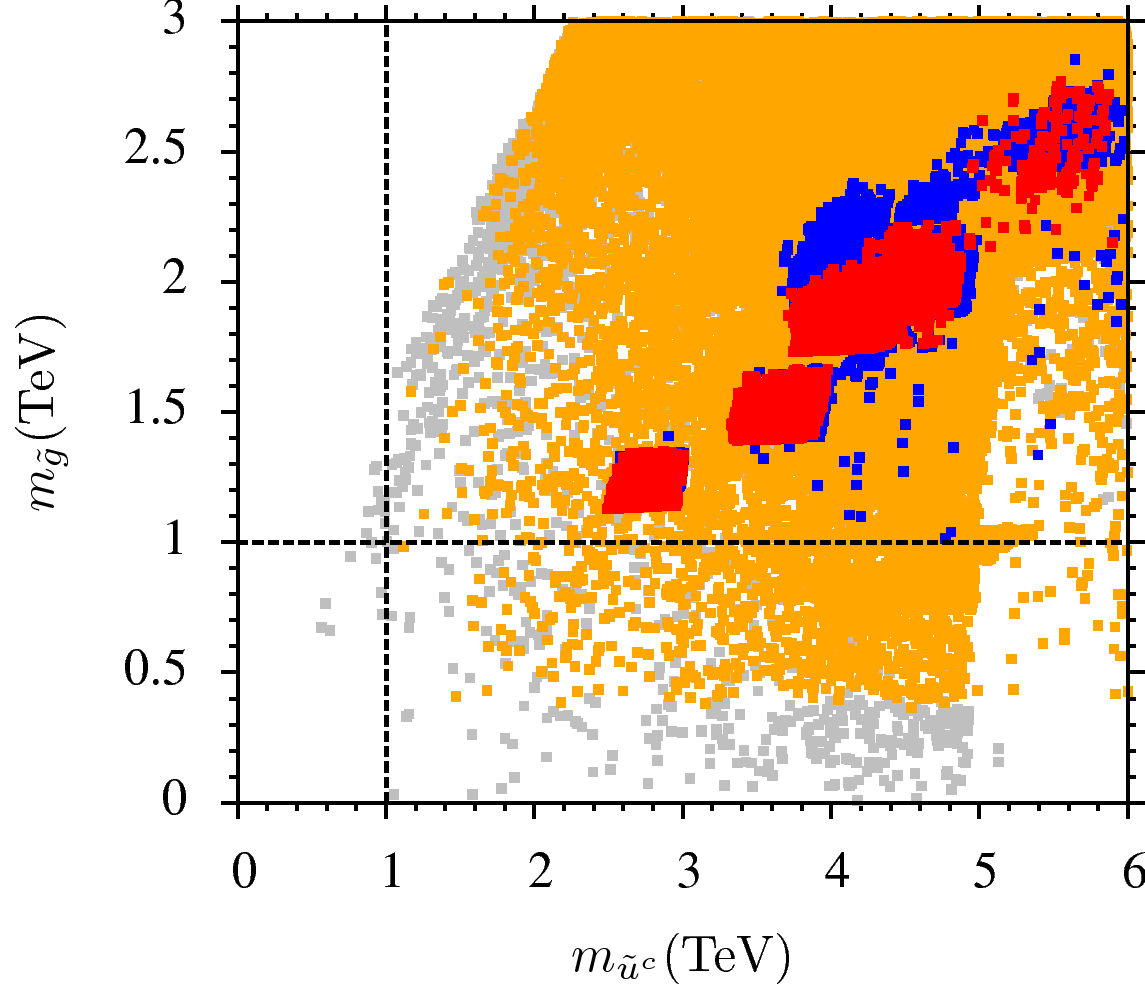}
}
\subfigure{
\includegraphics[totalheight=5.5cm,width=7.cm]{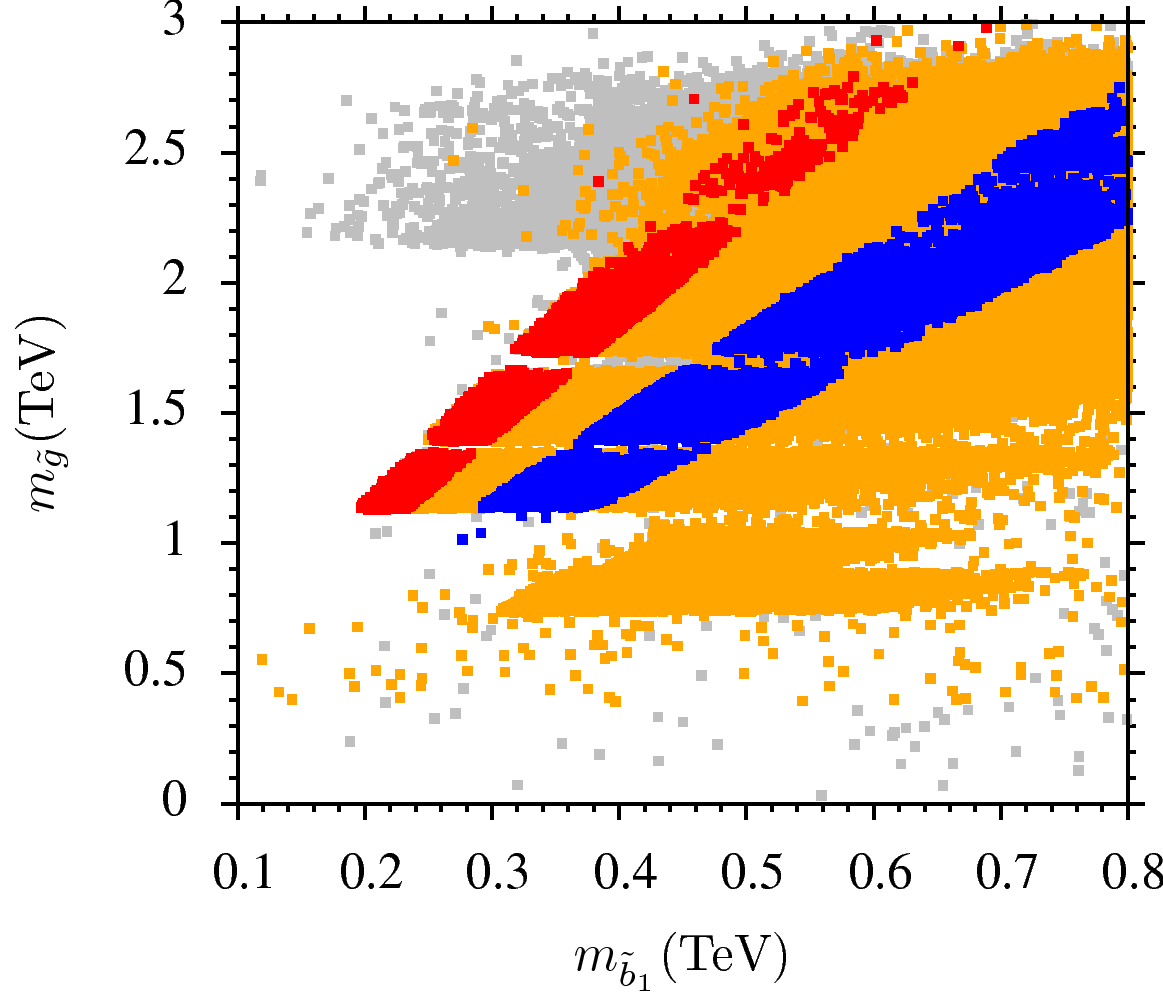}
}
\subfigure{
\includegraphics[totalheight=5.5cm,width=7.cm]{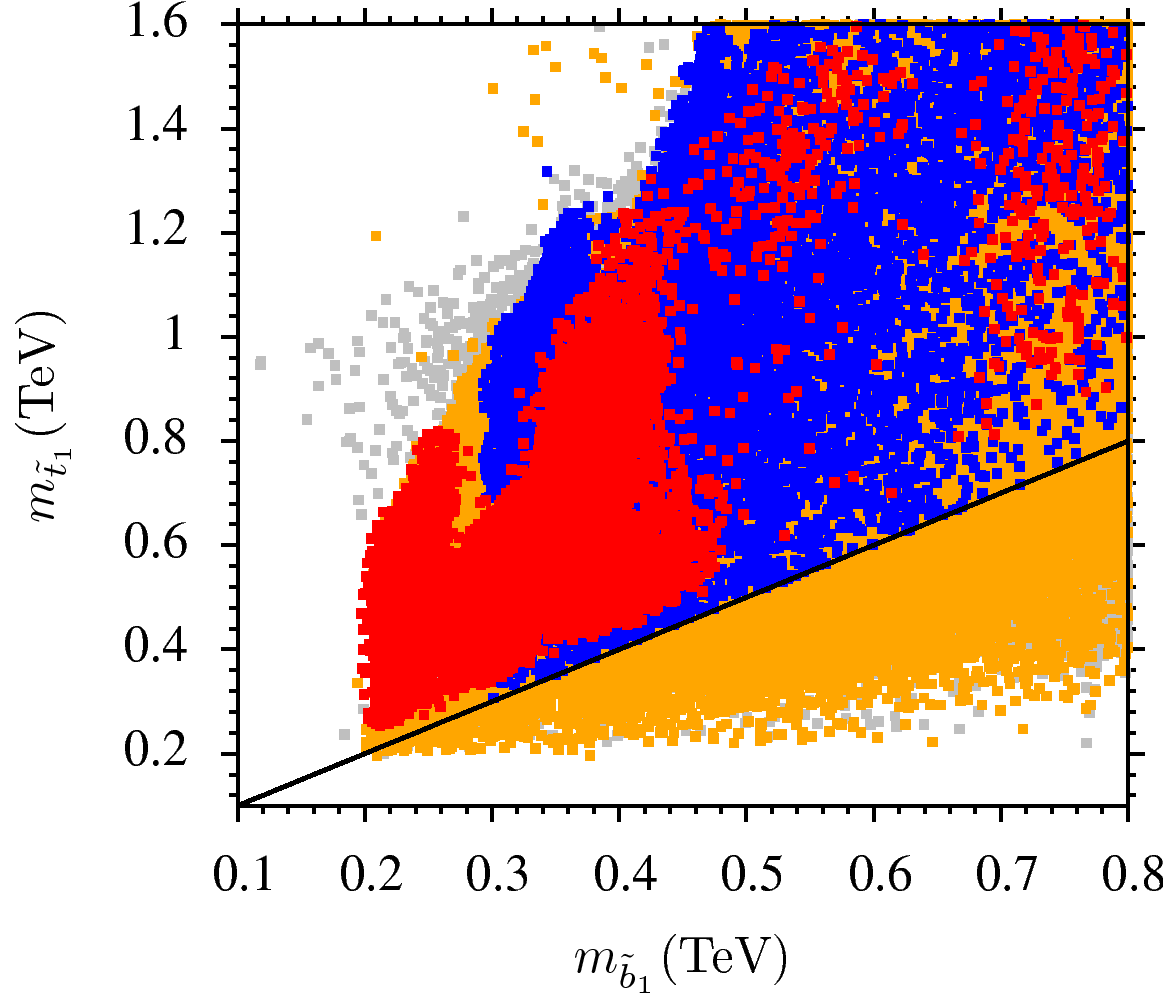}
}
\subfigure{
\includegraphics[totalheight=5.5cm,width=7.cm]{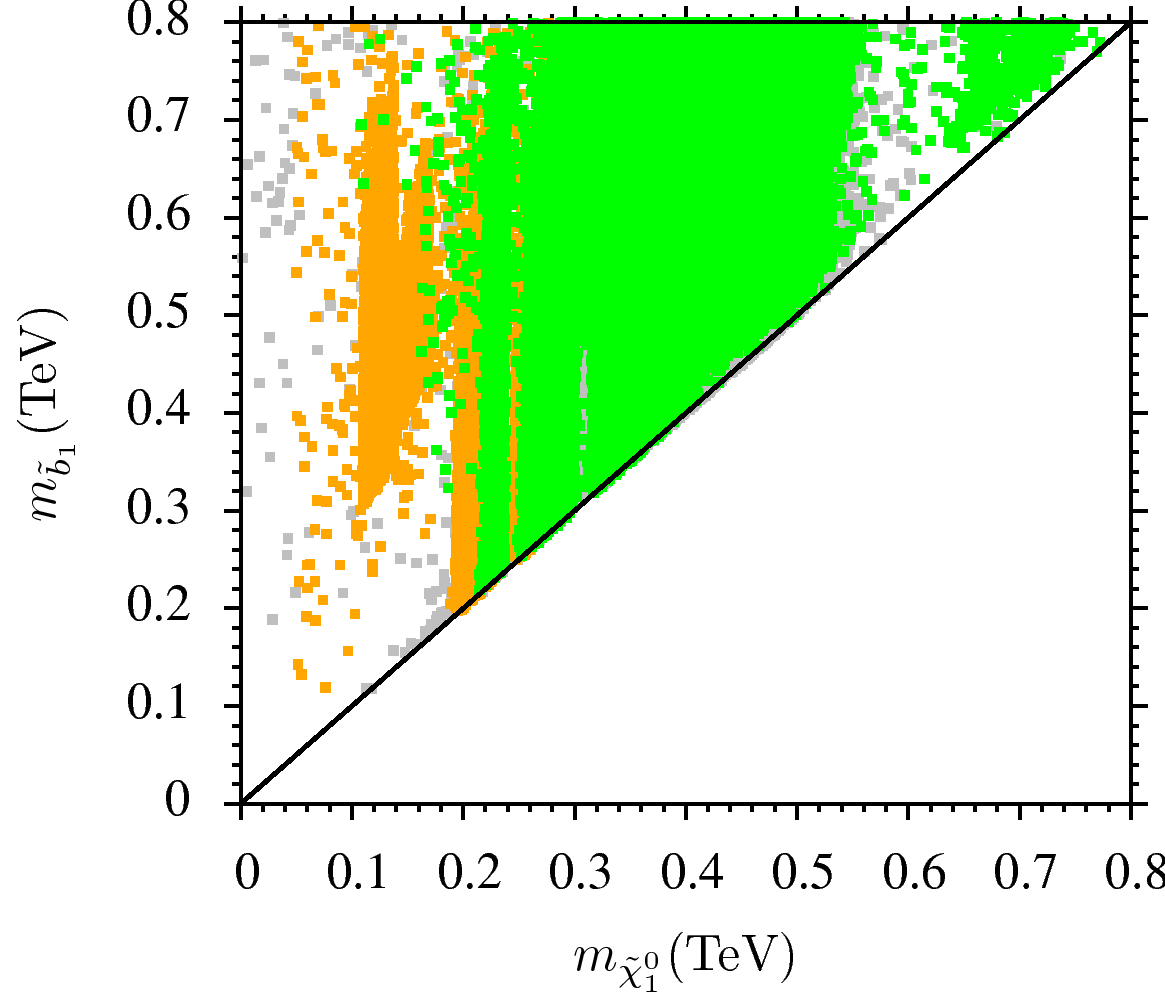}
}
\subfigure{
\includegraphics[totalheight=5.5cm,width=7.cm]{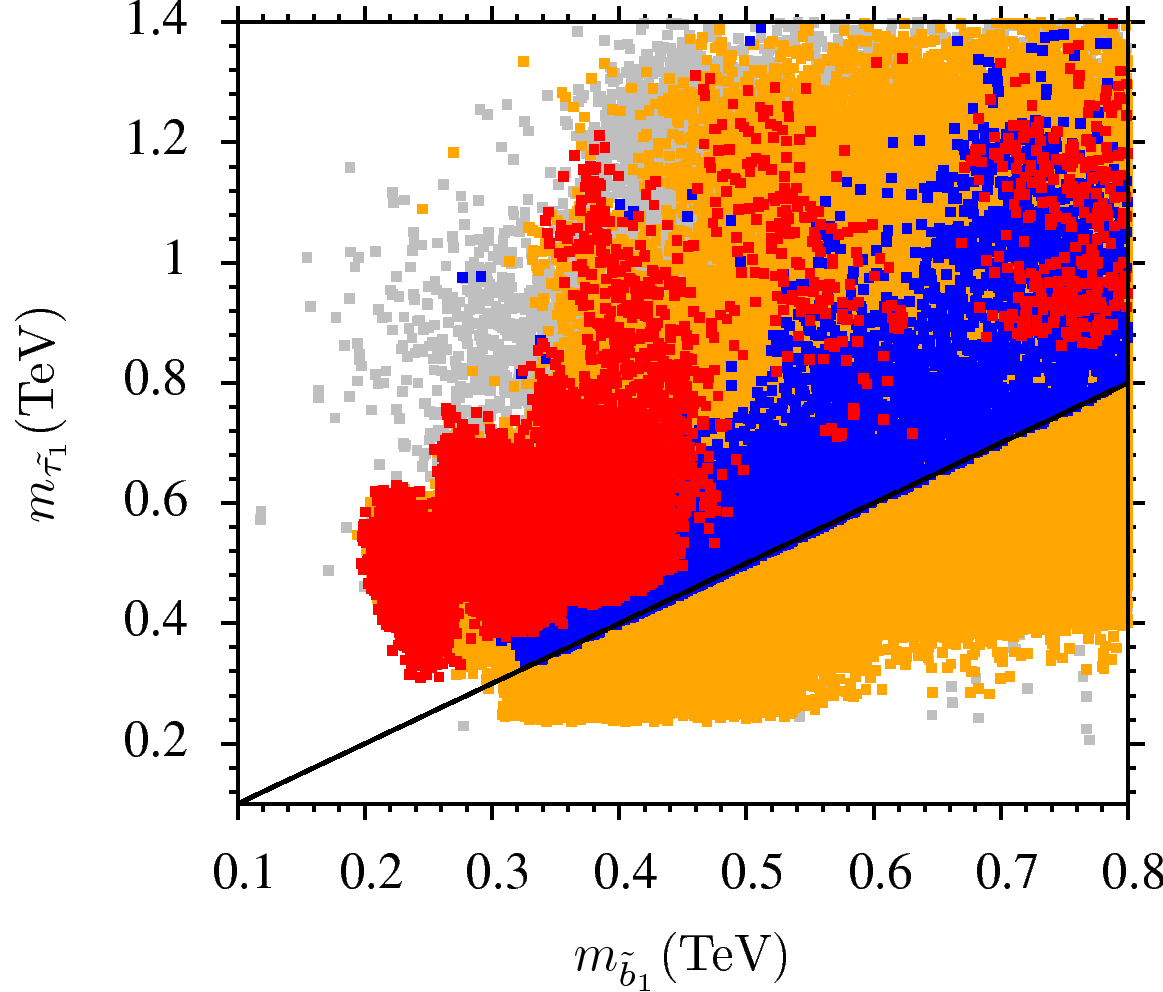}
}
\caption{Plots in the $m_{\tilde g}$ - $m_{\tilde {d}^c}$, $m_{\tilde g}$ - $m_{\tilde {u}^c}$,
 $m_{\tilde g}$ - $m_{\tilde {b}_1}$, $m_{\tilde {t}_1}$-$m_{\tilde {b}_1}$,
and $m_{\tilde {\tau}_1}$-$m_{\tilde {b}_1}$ planes.
Color coding is the same as in Figure~\ref{Fig1} except that green points in $m_{\tilde {t}_1}$-$m_{\tilde {\chi}^0_1}$ plane form a subset of orange points
which satisfies all current experimental bounds \cite{Aad:2011ib}.
}
\label{Fig3}
\end{figure}

In Figure~\ref{Fig3}, we show plots in
the $m_{\tilde g}$ - $m_{\tilde {d}^c}$,
$m_{\tilde g}$ - $m_{\tilde {u}^c}$,
 $m_{\tilde g}$ - $m_{\tilde {b}_1}$,
  $m_{\tilde {t}_1}$-$m_{\tilde {b}_1}$,
  $m_{\tilde {\tau}_1}$-$m_{\tilde {b}_1}$,
  and
  $m_{\tilde {t}_1}$-$m_{\tilde {\chi}^0_1}$,
   planes.
The color coding is
the same as in Figure \ref{Fig1}, except that the green points in the  $m_{\tilde{t}_1}$ - $m_{\tilde {\chi}^0_1}$ plane form a subset of orange points
and satisfy all current experimental bounds \cite{Aad:2011ib}.

From the results recently presented by the ATLAS \cite{Aad:2011ib} and CMS \cite{Chatrchyan:2011zy} collaborations, our findings are just beginning
to be tested.
In the $m_{\tilde {d}^c}$ - $m_{\tilde g}$ and $m_{\tilde {u}^c}$ - $m_{\tilde g}$  planes we show graphs for the first
generation squarks versus gluino mass.
From the $m_{\tilde {d}^c}$ - $m_{\tilde g}$ plane one sees that the
minimum value of $m_{\tilde {d}^c}$ in the neutralino-sbottom coannihilation scenario is around $800\,{\rm GeV}$, corresponding to a gluino mass of
$1.2 \,{\rm TeV}$.
From the plot in $m_{\tilde {u}^c}$ - $m_{\tilde g}$ plane, we see that
$m_{\tilde {u}^c}$ is heavy ($\gtrsim 2.2\, {\rm TeV}$), and lies well above the current experimental bound.

The plot in $m_{\tilde {b}_1}$-$m_{\tilde g}$ plane shows that the neutralino-sbottom coannihilation solutions (red points)
allow
a relatively light sbottom quark, with mass as low as $210 \,{\rm GeV}$. The corresponding gluino mass is around $1.2 \,{\rm TeV}$.
For gluino mass around $3\,{\rm TeV}$, the corresponding value of $m_{{\tilde b}_{1}}$ in this scenario is around $700 \,{\rm GeV}$.
The plot in $m_{\tilde {b}_1}$ - $m_{\tilde {t}_1}$
plane shows that in neutralino-sbottom coannihilation solutions, the minimum (maximum) value of the lighter stop quark mass is
of order $300\, {\rm GeV}$ ( $1.4\, {\rm TeV}$.)

In Figure~\ref{Fig3} we also show plots in $m_{\tilde {b}_1}$ - $m_{{\tilde \tau}_1}$ and  $m_{\tilde {b}_1}$ - $m_{{\tilde \chi}_1^0}$  planes.
In the $m_{\tilde {b}_1}$ - $m_{{\tilde \chi}_1^0}$ plane, the green points represent solutions that satisfy all the constraint mentioned above.
Points within 20\% of the line with unit slope represent {neutralino-sbottom} coannihilation, while points in blue color represent just NLSP sbottom solutions (without coannihilation.) From the graph we see that the minimum value of neutralino mass in either case
close to $210\,{\rm GeV}$. The minimum value of stau mass corresponding to the coannihilation scenario is about $500\,{\rm GeV}$.

\begin{figure}
\centering
\subfiguretopcaptrue
\subfigure{
\includegraphics[width=7.2cm]{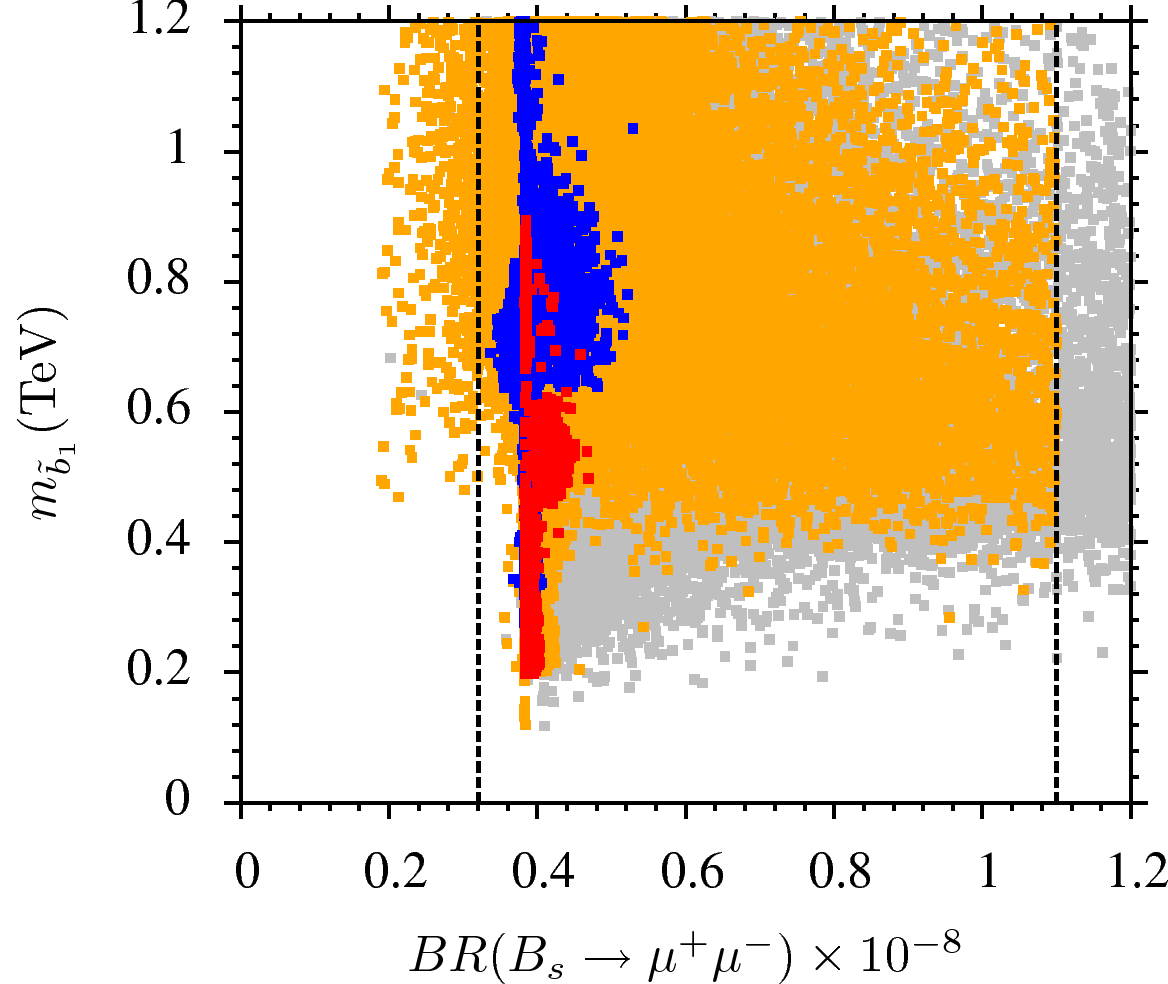}
}
\caption{Plot in $m_{\tilde {b}_1}$-$BR(B_{s} \rightarrow \mu^+ \mu^-)$ plane. Color coding is the same as in Figure \ref{Fig1}. Left vertical dashed line
corresponds to the SM prediction for $BR(B_{s} \rightarrow \mu^+ \mu^-)$ and right dashed line shows the current LHCd+CMS limit.
}
\label{Fig4}
\end{figure}

In Figure ~\ref{Fig4} we show a plot in the $m_{\tilde {b}_1}$ - $BR(B_{s} \rightarrow \mu^+ \mu^-)$ plane, with the color coding the same as in Fig.~\ref{Fig1}.
The vertical dashed line
on the left marks the SM prediction for $BR(B_{s} \rightarrow \mu^+ \mu^-)$, while the dashed line on the right shows the current LHCd+CMS limit.
As previously mentioned in section \ref{constraintsSection}, we have applied the latest combined LHCd+CMS limit
for $BR(B_{s} \rightarrow \mu^+ \mu^-)$ of $1.1 \times 10^{-8}$. This can be seen as a sharp cutoff in solutions represented by yellow points.
We observe that the neutralino-sbottom coannihilation solutions predict a value for $BR(B_{s} \rightarrow \mu^+ \mu^-)$ of around $4 \times 10^{-9}$, close
to the SM prediction of  around $3.2 \times 10^{-9}$. This prediction of a slight excess in
$BR(B_{s} \rightarrow \mu^+ \mu^-)$ for the neutralino-sbottom coannihilation scenario should be tested soon.

\begin{table}[h!]
\centering
\scalebox{0.8}{
\begin{tabular}{lccc}
\hline
\hline
                 & Point 1 & Point 2 & Point 3       \\
\hline
$M_{1/2}$          & 453   & 591   & 1544      \\
$m_{10}$           & 2394  & 3626  & 7972         \\
$m_{\overline{5}}$ & 519   & 577  &  926        \\
$\tan\beta$       &  10    & 12    & 10       \\
$A_{t}$           & -4347  & -7012  & -16000             \\
$A_{b}$           & 10080  & 9931  & 30010            \\
$m_{H_d}$         & 835    & 338  &  1121           \\
$m_{H_u}$         & 3065   & 4213  & 9116            \\
$sign(\mu)$        & +      & +   & +        \\

\hline
$m_h$            &123 & 125  & 127    \\
$m_H$            &536 & 1785 & 4251        \\
$m_{A}$          &533 & 1773 & 4223       \\
$m_{H^{\pm}}$    &541 & 1787 & 4252       \\

\hline
$m_{\tilde{\chi}^0_{1,2}}$
                 &194, 374  &260, 505     &703, 1330        \\
$m_{\tilde{\chi}^0_{3,4}}$
                 &822, 831 &2049, 2051    &4872, 4873    \\

$m_{\tilde{\chi}^{\pm}_{1,2}}$
                 &374, 840  &505, 2066  &1332, 4903   \\
$m_{\tilde{g}}$  &1135     & 1460     &3524    \\

\hline $m_{ \tilde{u}_{L,R}}$
                 &2546, 2611  &3785, 3875  &8429, 8597    \\
$m_{\tilde{t}_{1,2}}$
                 &372, 1688  &709, 2584  &1721, 5743     \\
\hline $m_{ \tilde{d}_{L,R}}$
                 &2548, 909 &3786, 1033 &8430, 2445    \\
$m_{\tilde{b}_{1,2}}$
                 &213, 1692 &414, 2610  &823, 5825    \\
\hline
$m_{\tilde{\nu}_{1}}$
                    & 729   & 905       & 1840         \\

$m_{\tilde{\nu}_{3}}$
                 &    472    & 579      &  748        \\
\hline
$m_{ \tilde{e}_{L,R}}$
                &739, 2308   &926, 3512 &1890,7745    \\
$m_{\tilde{\tau}_{1,2}}$
                &499, 2169    &622, 3371  &917, 7374    \\
\hline

$\sigma_{SI}({\rm pb})$
                &6.23 $\times 10^{-11}$ & 7.61$\times 10^{-12}$ & 5.18$\times 10^{-13}$ \\

$\sigma_{SD}({\rm pb})$
                &2.92 $\times 10^{-7}$ & 1.87$\times 10^{-8}$ & 5.73$\times 10^{-10}$ \\

$\Omega_{CDM}h^2$
                & 0.08  & 6.8    &  0.09     \\

\hline
\hline
\end{tabular}
}
\caption{ Point 1 shows the minimum value of sbottom mass ($ 213\,{\rm GeV}$) for neutralino-sbottom coannihilation.
Point 2 represents a solution with sbottom NLSP ($414\,{\rm GeV}$), more than 50\% heavier than the LSP neutralino.
Point 3 represents a neutralino-sbottom coannihilation solution with a heavier sbottom mass ($823\,{\rm GeV}$).
}
\label{table1}
\end{table}

In Table~\ref{table1} we present three characteristic benchmark points which satisfy all constraints presented in Chapter 3.
Points 1 and 3 respectively represent the  minimum and maximum values of the sbottom mass ($\sim 210$ GeV and $\sim 820\,{\rm GeV}$),
corresponding to neutralino-sbottom coannihilation. Point 2 represents just an NLSP sbottom solution (without insisting on correct dark matter relic abundance value), with the stop and stau masses relatively close to it.
Since the LSP is essentially a pure bino, both its spin-independent ($\sim 10^{-13}-10^{-11} \rm{pb}$) and 
spin-dependent ($\sim 10^{-10}-10^{-7} \rm{pb}$) cross sections on nucleons are
rather  small \cite{Gogoladze:2010ch}.
 Consequently it won't be easy  to detect the LSP in direct and indirect experiments. However, as shown in \cite{Adeel} and \cite{Feldman}, the LHC
supersymmetry searches can be exploited to probe cross sections of this magnitude.

\section{Conclusions\label{conclusions}}

We have described in detail the conditions under which the neutralino-sbottom coannihilation scenario can be realized in
supersymmetric $SU(5)$. In particular we have identified, for the first time we believe, the minimum number of soft supersymmetry breaking parameters
that are required in order to have NLSP sbottom and neutralino-sbottom coannihilation in $SU(5)$. The coannihilation scenario predicts the
existence of relatively heavy ($\gtrsim 1\, {\rm TeV}$) gluino and first two generation squark masses. The NLSP sbottom in the coannihilation scenario is
quasi-degenerate in mass with the LSP neutralino, and it can be as light as $210\, {\rm GeV}$ or so, without running into conflict with the current
supersymmetry search at the LHC. We also highlight a few benchmark points which can be tested at the LHC.

\section*{Acknowledgments}
We thank  M.~Adeel Ajaib, T.~Li, M.~ Papucci and O.~ Gonzalez Lopez for valuable discussions.
This work is supported in part by the DOE Grant No. DE-FG02-91ER40626
(I.G., S.R. and Q.S.). This work used the Extreme Science
and Engineering Discovery Environment (XSEDE), which is supported by the National Science
Foundation grant number OCI-1053575.

\end{document}